\documentclass[conference]{IEEEtran}
\IEEEoverridecommandlockouts
\usepackage{cite}
\usepackage{amsmath,amssymb,amsfonts}
\usepackage{algorithmic}
\usepackage{subfigure}
\usepackage{makecell}
\usepackage{color}
\usepackage{graphicx}
\usepackage{textcomp}
\usepackage{xcolor}
\usepackage{multicol}
\usepackage{multirow}
\usepackage{ulem}

\usepackage{setspace}
\def\BibTeX{{\rm B\kern-.05em{\sc i\kern-.025em b}\kern-.08em
    T\kern-.1667em\lower.7ex\hbox{E}\kern-.125emX}}
\begin{document}

\title{Real Time Lateral Movement Detection based on Evidence Reasoning Network for Edge Computing Environment}

\author{
\IEEEauthorblockN{Zhihong Tian\IEEEauthorrefmark{2},  Wei Shi\IEEEauthorrefmark{3}, Yuhang Wang\IEEEauthorrefmark{2}, Chunsheng Zhu\IEEEauthorrefmark{4}, Xiaojiang Du\IEEEauthorrefmark{5}, Shen Su\IEEEauthorrefmark{1}\IEEEauthorrefmark{2}, Yanbin Sun\IEEEauthorrefmark{2}, Nadra Guizani\IEEEauthorrefmark{6}}
\IEEEauthorblockA{\IEEEauthorrefmark{1}Corresponding Author johnsuhit@gmail.com}
\IEEEauthorblockA{\IEEEauthorrefmark{2}Cyber space Institute of Advanced Technology, Guangzhou University, Guangzhou, China}
\IEEEauthorblockA{\IEEEauthorrefmark{3}School of Information Technology Faculty of Engineering and Design, Carleton University, Ottawa, Canada}
\IEEEauthorblockA{\IEEEauthorrefmark{4}Department of Electrical and Computer Engineering, \\ The University of British Columbia, Vancouver, British Columbia, Canada}
\IEEEauthorblockA{\IEEEauthorrefmark{5}Department of Computer and Information Sciences, Temple University, Philadelphia, USA}
\IEEEauthorblockA{\IEEEauthorrefmark{6}ECE Department, Purdue University, Indiana, USA}
}

\maketitle

\begin{abstract}
Edge computing is providing higher class intelligent service
and computing capabilities at the edge of the network. The aim is to ease
the backhaul impacts and offer an improved user experience, however,
the edge artificial intelligence exacerbates the security of the cloud computing
environment due to the dissociation of data, access control and
service stages. In order to prevent users from using the edge-cloud computing
environment to carry out lateral movement attacks, we proposed a
method named CloudSEC meaning real time lateral movement detection
based on evidence reasoning network for the edge-cloud environment.
The concept of vulnerability correlation is introduced. Based on the
vulnerability knowledge and environmental information of the network
system, the evidence reasoning network is constructed, and the lateral
movement reasoning ability provided by the evidence reasoning network
is used. CloudSEC realizes the reconfiguration of the efficient real-time
attack process. The experiment shows that the results are complete and
credible.
\end{abstract}


\section{Introduction}

Cloud computing services are critical information platform, application and infrastructures resources that users access via Internet[1-3]. These services, offered by companies such as Google and Amazon, enable customers to leverage powerful computing resources that would otherwise be beyond their means to purchase and maintenance. The success of the cloud computing has already changed the appearance of the world information infrastructure.

In order to ease the backhaul impacts and offer an improved user experience, edge computing are designed to share the load of cloud center, provides an IT service environment and computing capabilities at the edge of the network[4,5]. The environment of edge computing is characterized by low latency, proximity, high bandwidth. Combined with the traditional cloud computing environment, the edge computing enables innovative services such as e-Health, augmented reality, smart camera, gaming and industry automation \cite{edge}.

The "edge-cloud" model extends the complexity of cloud services [7-11]. More authorities have to be transferred to the edge participants, and more stages and data interactions will be involved during the procedure of services [12-15]. As the result, it turns out to be impractical to prevent all intrusions. Thus, monitoring the operation of a system is essential to its security. To that end, researchers collect monitoring information to estimate the current state of the system based on the usage patterns, detect whether there exist intrusions, and drive response actions.


Lateral movement techniques are frequently used to launch cyber-attack, especially in the hierarchical architecture system. In the edge-cloud environment, two main challenges are involved in the lateral movement detection.

1. The data persistence in edge computing may be different with the traditional scenario, there is less certainty in where data originated from, whether the data would be persistence, and where it will be stored. Since the edge nodes are often the limited memory and computationally constrained devices. The traditional lateral movement detection methods which need significant manual effort and business correlated knowledges lacks the opportunity to flourish in such scenario.

2. The underlying architecture of the edge computing environment is often dynamic, for example, the vehicular frog computing system based on the vehicle ad hoc networks. As the result, the lateral movement detection methods that rely on detecting changes in nodes behaviors may not be applied successfully on the edge environment.

Existing solutions are with few considerations for such scenario.  When a host is discovered to be compromised, there are a couple of fundamental questions that should always be asked: What was the route this attack traverse? How was the movement possible? What was the end target? What controls were executed to make the threat persistent? In order to meet these forensic requirement, the lateral movement detection method should be applied to the edge-cloud computing environments.

In this paper, we present a lateral movement detection method based on Evidence Reasoning Network (ERN) for edge-cloud computing environment referred to as CloudSEC. CloudSEC introduces the concept of vulnerability correlation and built an ERN based on the network system vulnerabilities and environmental information. Experimental results show that CloudSEC supplies a complete and credible evidence chain and it also has the capacity of real-time lateral movement reasoning. These provide a strong guarantee for the rapid and effective evidence investigation. The contributions of CloudSEC are threefold: First, it offers a list of concrete evidence on the detected attack which gives more confidence to the cloud service provider. Second, it enables the cloud service provider to be fully aware the consequences of an attack. Finally, it enables the cloud service provider to better determine an appropriate course of actions that needs to be taken.

The remainder of this paper is organized as follows: In section 2, we present related previous works, such as event correlation in virtual machine. In Section 3, we depict the overall structure of our method. In Section 4, we present the details of our proposed lateral movement detection method and its correlation algorithms. The experiments conducted to evaluate our method are discussed in Section 5. Section 6 concludes the paper with a summary and discussion of future research directions.

\section{Related Works}

Up to today, there have been several proposed techniques of analyzing android malware cloud data [16-20]. In this paper, we focus exclusively on the related works using forensics techniques on monitoring virtual machines in cloud computing environment.

Most mature forensic investigation tools such as EnCase [21] and Safeback [22] focus on capturing and analyzing evidence from media stores on a single host. Mnemosyne is a dynamically configurable advanced packet capturing application that supports multi-stream capturing, sliding-window based logging to conserve space, and query support on collected packets [23]. Evidence graph network forensic analysis mechanism [24][25] includes effective evidence presentation, manipulation and automated reasoning. Although it is nice to present evidence correlation in graphic mode, this system is still a prototype and lacks the effective capability of inference. Tian et al. [26][27] developed a network intrusion forensics system based on transductive scheme and Dempster-Shafer Theory (DST) that can detect and analyze efficiently computer crimes in networked environments, and extract digital evidence automatically. ForNet [28] is a novel distributed logging mechanism that focuses on network forensic evidence collection rather than evidence analysis.
These approaches rely on the long term log data collection or on the statistical-based methods are not suitable for the poor data persistence environment.

Different from the above-mentioned research, Walls et al. [29] developed DEC0DE, a system for recovering information from phones with unknown storage formats, which was a critical hurdle in forensic triage. Because phones have myriad of custom hardware and software, and all stored data must be examined. Via flexible descriptions of typical data structures, and using a classic dynamic programming algorithm, DEC0DE system is able to identify call logs and address book entries in phones across a wide range of models and manufacturers. In [30], authors perform extensive study on existing fuzzy hashing algorithms with the goal of understanding their applicability in clustering similar malware. They developed a memory triage tool that uses fuzzy hashing to intuitively identify malware by detecting common pieces of malicious code found within a process. In [31], a host-based intrusion detection system gained a high degree of visibility as it is integrated into the host's monitoring process. A new approach, based on the k-Nearest Neighbor (kNN) classifier, is used to classify program behavior as normal or intrusive. Ko et al. [32] introduced an approach that integrates intrusion detection techniques with software wrapping technology to enhance a system's ability to defend against intrusions. In particular, they employ the NAI Labs Generic Software Wrapper Toolkit to implement all or part of an intrusion detection system as ID wrappers.

Because the above host-based methods operate at a user level, unfortunately, these systems are quite susceptible to attacks once an attacker has gained access privilege to a host. Besides, an operating system crash will generally cause the system to fail to open. Since the host-based method runs in the same fault domain as the rest of the kernel, this will often cause the entire system to crash or allow the attacker to compromise the kernel [33-35].

\section{The Proposed CloudSEC Framework}

\begin{figure}
\centering
\includegraphics[width=0.45\textwidth]{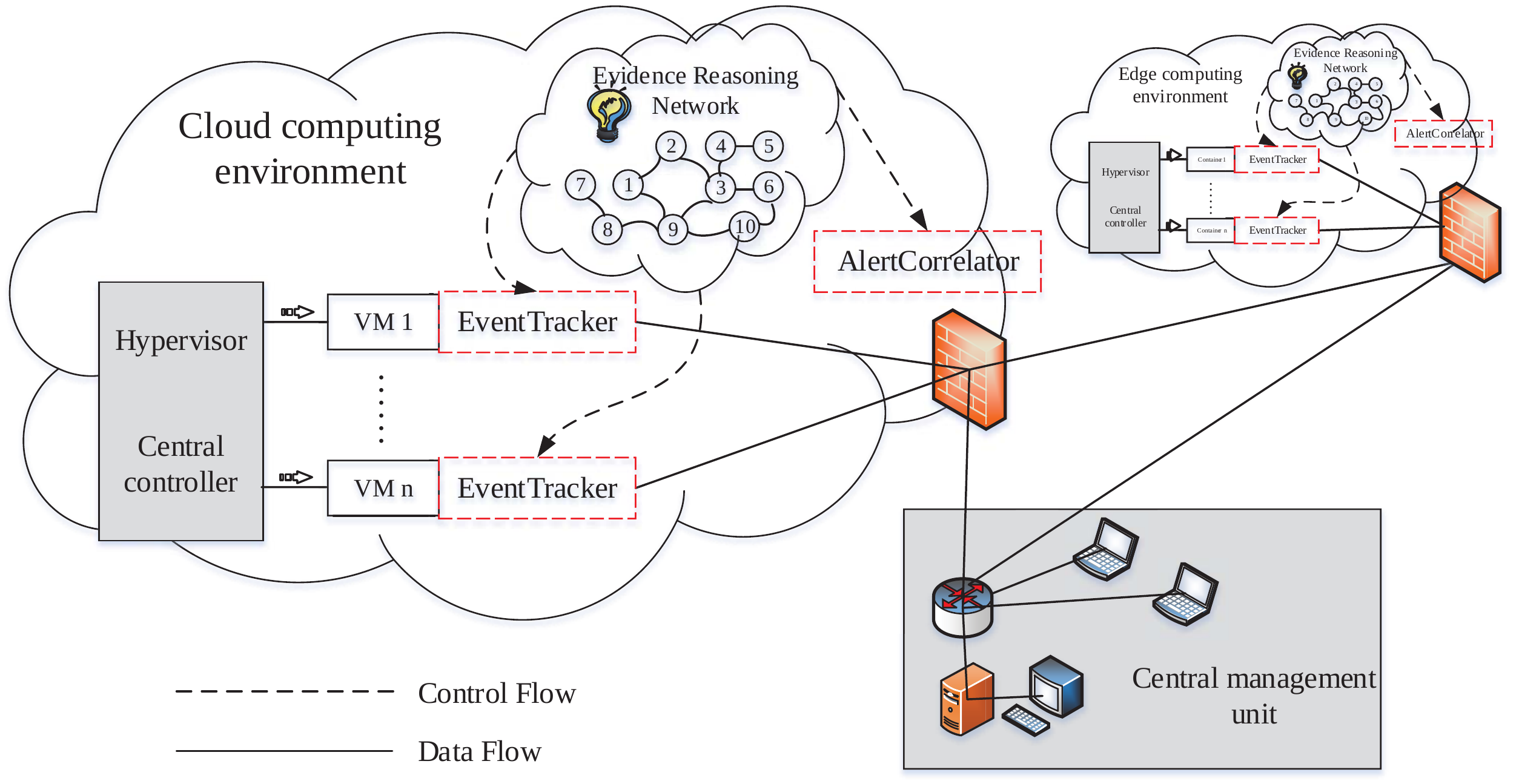}
\caption{The Architecture of CloudSEC} \label{fig1}
\end{figure}

The overview of the proposed CloudSEC architecture is shown in Figure~\ref{fig1}. CloudSEC consists of two components: EventTracker and AlertCorrelator respectively. Each Virtual Machine (VM) or the Container has a built-in EventTracker that used to monitor user activities and detect intrusions by auditing the event log on this system and analyzing commands and tracing system calls made by the users. AlertCorrelator is located at the edge of the cloud computing environment. It correlates and analyzes alerts generated by multiple distributed network Intrusion Detection Sensors (NIDS) deployed in a specific cloud computing environment. A central management unit is responsible for exchanging information, and put in place a set of evaluation criteria used to evaluate the trustworthiness of the results. In the following sections, we describe these two above-mentioned components in detail.

The main idea behind CloudSEC is the modeling of ERN. It is known that lateral movement is a sequence of small attack units/steps happened in different times and network locations following a certain logic. Each step of an attack can be considered as a preparation step for the next one. In ERN model, a detected attack step is referred to as an evidence. If we can discover all hidden correlation in an evidence chain, then a lateral movement is detected. Based on the concept of vulnerability and vulnerability correlation, an ERN is constructed to correlate and reasoning out evidence chains and eventually achieve the goal of lateral movement detection. To ease the understanding of ERN, we first describe the concept of vulnerability and vulnerability correlation.
\subsection{Vulnerability and Vulnerability Correlation}

The concept of vulnerability refers to the defects of the computer system following certain security strategy. Since the vulnerability is an intrinsic factor of security incidents, we define vulnerability and vulnerability correlation as the following:

Definition 1. (Vulnerability). A vulnerability is a defect exists in a software system or software component. The exploitation and utilization of such a defect would violate one or more security policies and adversely affect the confidentiality, identifiability, and usability of the software system. We note a vulnerability as v in the rest of this paper.

Definition 2. (Vulnerability Correlation). Supposing that a software system $S$ has $n$ vulnerabilities $V[S]=\{v_1,v_2,\cdots,v_n\}$, if $\exists V_{li}(i=1 \sim k \wedge k \le n) \in V[S]$ , so that the attacker could launch a multiple stages attack with $v_{l1},\cdots,v_{lk}$ , then we say $v_{l1},\cdots,v_{lk}$ are correlated, or we say there exists a correlation among $v_{l1},\cdots,v_{lk}$.

\begin{figure}
\centering
\includegraphics[width=0.25\textwidth]{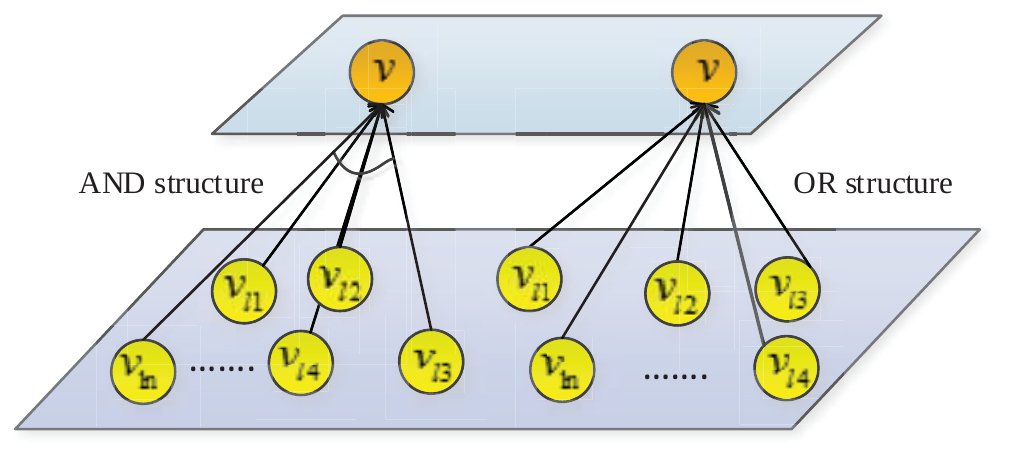}
\caption{Two basic structures of vulnerability correlation} \label{fig4}
\end{figure}

Vulnerability correlation can be represented either as an AND or an OR structure (as shown in Figure~\ref{fig4}).

In an AND structure, the premise of utilizing vulnerability v to attack the target system is to utilize all the vulnerabilities of $v_{li}(1 \leq i \leq n, n \geq 2)$. In an OR structure, the premise of utilizing vulnerability v to attack the target system is to utilize any one of the vulnerabilities in $v_{li}(1 \leq i \leq n, n \geq 1)$.

Given the nature of an attack is to exploit one or more existing vulnerabilities on a computer system, we need to associate the attacks to the vulnerabilities. To do so, we map the attack feature space into the vulnerability space, so that we could describe the attack correlations in the perspective of vulnerability correlations. Therefore, in order to form a credible evidence chain, we construct an ERN to correlate discrete attack evidences based on vulnerability correlation.

\subsection{Evidence Reasoning Network Model}

We use a directed graph to describe the connectivity and vulnerability correlations of an information system network, and denote it as an evidence reasoning network.

Definition 3. (Evidence Reasoning Network). Evidence reasoning network (ERN) is a directed graph G that is described using a set of 5-tuples\\ $\{N, E, L, W, D\}$:

\begin{itemize}
\item	$N$ is a set of n vertex, each of which carries some vulnerabilities information. A tuple $(n_i, v_i)$ together represents a network node, where $n_i$ is the ID of the network node and $v_i$ is a set of vulnerabilities on $n_i$ and $0 \le i \le n, n\ge 0$.

\item $E$ is a set of $e$ directed links, each of which is an ordered pair: $e_j(e_j \in E) = ((n_x, v_x), (n_y, v_y))$, where $n_x$ contains $v_x$ and $n_y$ contains $v_y$, $0 \le j \le e$, $0 \le x \le n$, $0 \le y \le n$. Such a link $e_j$ indicating that $(n_x, v_x)$ interconnects with $(n_y, v_y)$, and $v_x$ and $v_y$ have vulnerability correlation. We say $(n_x, v_x)$ is a parent node, and $(n_y, v_y)$ is a child node. For simplicity, we denote each such directed link $e_j$ as $n_x \to n_y$. For $\forall(n_i, v_i) \in N$, we define $IN(n_i, v_i)=\{E_I | E_I = ((n_l,v_l),(n_i,v_i)), E_I \subseteq E, (n_l, v_l) \in N, 0 \le l \le n\}$ as a set of incoming link with node $(n_i, v_i)$ as a child node. Similarly, we define $OUT(n_i, v_i)=\{E_J | E_J = ((n_i,v_i),(n_m,v_m)), E_J \subseteq E, (n_m, v_m) \in N, 0 \le m \le n\}$ as a set of directed links that have node $n_i$ as a parent node.

\item $L$ is a set of logical expressions representing the relationships among the directed links using AND and OR logical expression operators along with brackets. $L$ is one-to-one mapped to $N$, expressing the relationships among the directed links going in or out each specific node. $\forall (n_i, v_i)$, if $IN(n_i, v_i)= \{e_1,e_2,e_3 | e_1=((n_1, v_1), (n_i, v_i)), e_2=((n_2, v_2), (n_i, v_i)), e_3=((n_3, v_3), (n_i, v_i))\}$ and $l_i={(e_1 \wedge e_2) \vee e_3}$, where $l_i \subseteq L, n_l \in N, 0 \le l \le n \}$, Then parent nodes $(n_1, v_1)$ and $(n_2, v_2)$ both fall into an AND relationship with $(n_i, v_i)$. On the other hand, $(n_3, v_3)$ and $(n_i, v_i)$ fall into an OR relationship.

\item $W$ is defined as a set of risk weights for each vertex in $N$. $W$ has an one-to-one mapping to $N$. $\forall {n_i} \in N$, we define $w_i=(f_i + p_i + r_i)/3$, where $f_i \in [0,1]$ denotes the functional value of vertex $(n_i, v_i)$, including information servers, database servers, work stations, etc; $p_i \in [0,1]$ denotes the probability of successful exploitation of the vulnerability $v_i$ on vertex $(n_i, v_i)$; and $r_i \in [0,1]$ denotes the security impact of vulnerability $v_i$. $W$ is defined to setup standard for evaluating the generated evidence chain.

\item $D \in \{qn, pptr\}$ contains a set of data structures for each vertex in $N$. It has an one-to-one mapping to $N$. Here $qn$ is a circular queue of length $k$ for evidence storage. Each element of the queue may be represented as $qn=\{ts,cptr,wt\}$, here $ts$ is the timestamp; $cptr$ is the pointer pointing to the child node used to reconstruct an attacking process flow; $s$ is used to show the state of this node. Its value ranges from 0 to 3, representing respectively: a start node, both a start node and a virtual node, an intermediate node, or a virtual node; $wt$ is the risk weight. $pptr$ stores a set of pointers pointing to the parent nodes of this vertex. Such reversed pointers (i.e. pointers pointing to the parent nodes) are used to speed up the evidence reasoning process. For simplicity, we use `.' to index each specific element. For example, $d_i.pptr$ or $d_i.qn_i.ts$.
\end{itemize}

The following set of steps describe briefly the procedures to generate an ERN:

\begin{enumerate}	
\item Report the topology of the target network by device users such as the administrators or the node hosts of the target network to be protect;

\item Use existing vulnerability scanning tools such as Nmap and Nessus to probe the target network from different locations within and outside the managed network domain;

\item	Construct the vertex set;

\item	Traverse all vertex in the target network and construct recursively a set of directed links $E$ and the logical expression set $L$ based on node connectivity and vulnerability correlations;

\item	Compute $W$ using the predetermined risk weights and initialize data structure set $D$.

\end{enumerate}

Obviously, the execution of step 1 and 2 depend on the characteristics and scale of the target network topology. However the efficiency and executability of the forensics can be solved by pre-computing the ERN a priori.

\begin{figure}
\centering
\includegraphics[width=0.28\textwidth]{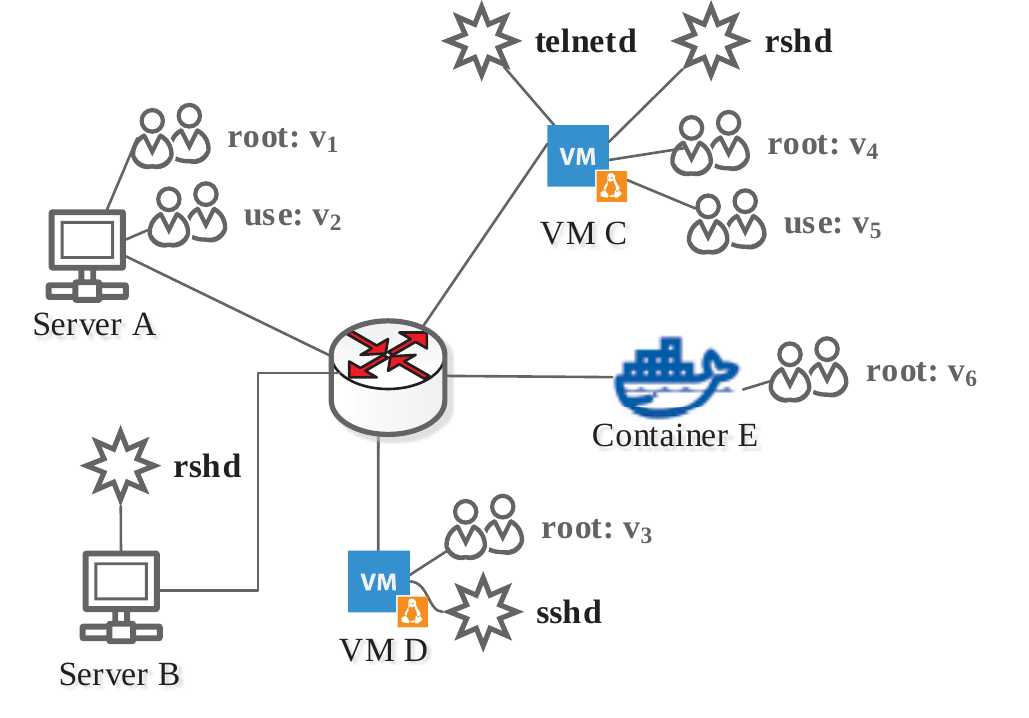}
\caption{Example of information system network} \label{fig5}
\end{figure}

Figure~\ref{fig5} illustrates an example of a network information system including 5 linux systems ($A$, $B$, $C$, $D$, and $E$). Server $A$ has vulnerability $v_1$ that threatens the root users, and vulnerability $v_2$ that threatens regular users. Service \emph{rshd} is running on $B$, which permits root users of $A$ and $D$ to gain remote access on $B$ executing shell command. The VM $C$ runs telnetd and rshd services. \emph{rshd} allows only root user of VM $D$ to execute shell command remotely on $C$. $C$ also has vulnerability $v_4$ that threatens all root users and vulnerability $v_5$ that threatens regular users. $D$ runs sshd service. It has vulnerability $v_3$ that threatens root users. The edge container $E$ has vulnerability $v_6$ that threatens all root users. All regular users of $A$ and the root user of $C$ could gain remote access to $D$ as root users via ssh. The root users of $B$ and $E$ could gain remote access to $C$ as a root user through telnet.

\begin{figure}
\centering
\includegraphics[width=0.2\textwidth]{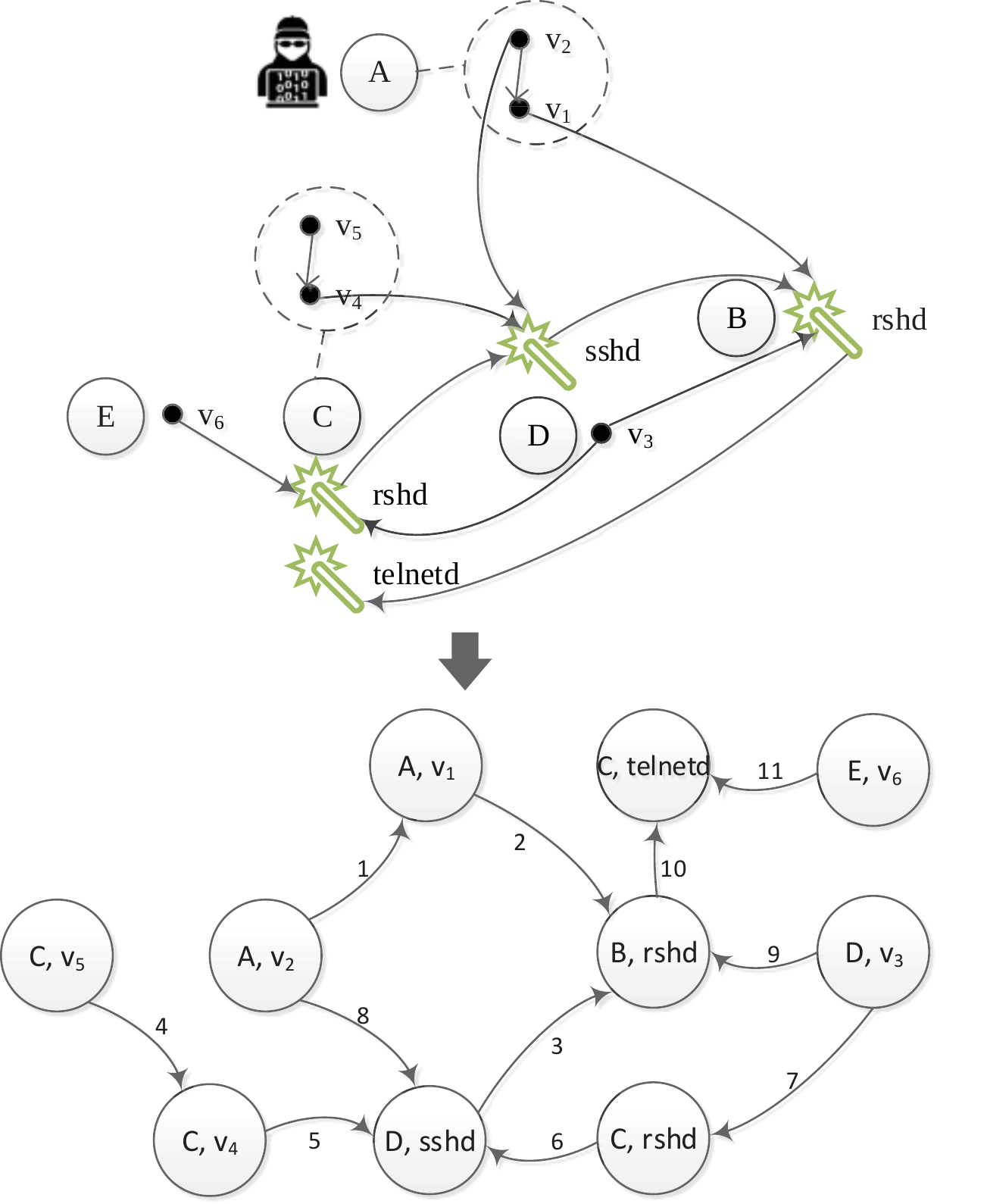}
\caption{Example of evidence reasoning network} \label{fig:ERNExample}
\end{figure}

The constructed ERN of the above information system network is shown in Figure~\ref{fig:ERNExample}, including 10 network nodes, and 11 directed links.

\subsection{Evidence Chain Reasoning}

ERN provides a framework for the reconstruction of attacking flow and evidence chain reasoning. Each evidence chain can be represented using a subgraph of  ERN. The main purpose of evidence preprocessing is to map the extracted evidence out of the attack flows to the vertex of ERN, noted as function $map()$.

Evidence chain reasoning could be abstracted as the following: assume there is a time sequenced evidence list: $\varepsilon_1,\varepsilon_2,\cdots,\varepsilon_n$, for $\forall \varepsilon_i=1,2,\cdots,n$, searching for the evidence set within the range of $\varepsilon_1,\varepsilon_2,\cdots,\varepsilon_{i-1}$, such that there exists a directed link from  $map(\varepsilon_h)$ to $map(\varepsilon_i)$ in the ERN, noted as $ map(\varepsilon_h) \to map(\varepsilon_i)$, $0 \le h \le n$.

The procedure of evidence chain reasoning includes the following 3 steps:

Step 1. Initialization: fetch an evidence $\varepsilon_i$, and map it to a vertex in the ERN of a given system, e.g., $map(\varepsilon_i)=n_i$. Then set $d_i.qn_i.ts$ as the timestamp $t_i$;

Step 2. Association analysis: the analysis methods depend on the number of links pointing to a vertex $n_i$ in ERN:

\begin{enumerate}
\item When $IN(n_i)=0$, $\varepsilon_i$ is the start point of this ERN. Then set $d_i.qn_i.s=0$, and $d_i.qn_i.wt=w_i$;

\item When $IN(n_i)>0$, use reverse index data structure $d_i.pptr$ to find all the parent nodes of $n_i$. If a parent node $n_j$'s $d_j.qn$ is non-empty, we say $(n_j,n_i)$ is a truth-value link.

\begin{itemize}
\item If the logical expression $l_i$ returns a true value with all the above-mentioned truth-value links, then we say $\varepsilon_i$ is an intermedia node. Thus we set $cptr=d_i.qn_i$ for all parent nodes that make $l_i$ true, and set $d_i,qn_i.s=2$, $d_i.qn_i.wt=w_i$;

\item If the logical expression $l_i$ returns a false value with all the above-mentioned truth-value links, then we know that $\varepsilon_i$ has no correlation with the other evidences. Thus we consider $\varepsilon_i$ as a start node, and set $d_i.qn_i.s=0$, $d_i.qn_i.wt=w_i$ accordingly.

\end{itemize}
\end{enumerate}

Step 3. Evidence chain generation: conduct breadth-first search from a vertex $n_i$ in the ERN, of which the in-degree is 0 (i.e. $IN(n_i)=0$), and generate evidence chain base on the $cptr$ of the traversed vertices.

\begin{figure}
\centering
    \subfigure[Evidence chain reasoning process]{
		\label{fig:ECRP}
		\includegraphics[width=0.2\textwidth]{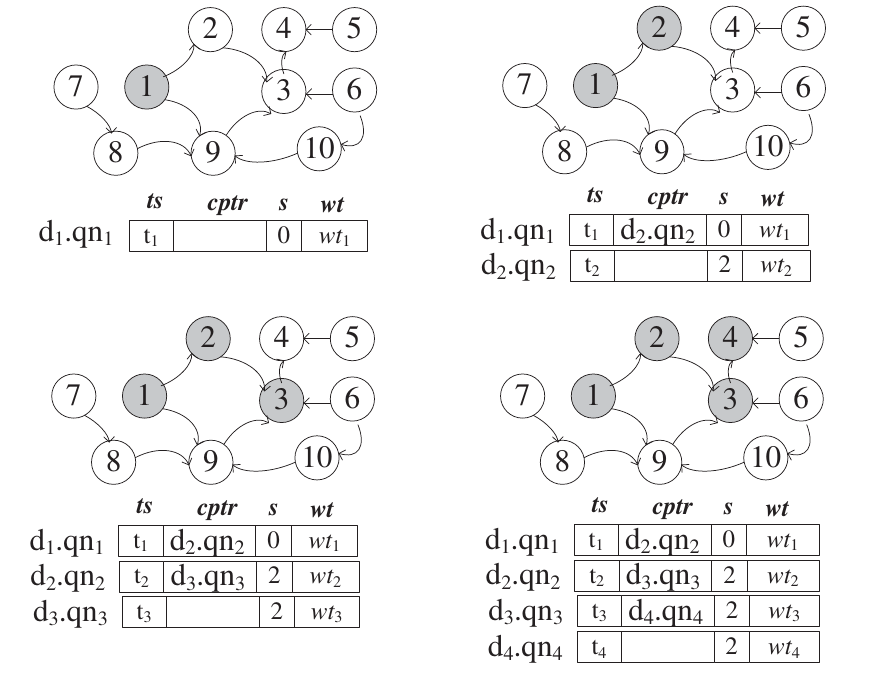}
	}
	\subfigure[Timing independent evidence chain reasoning process]{
		\label{fig8}
		\includegraphics[width=0.2\textwidth]{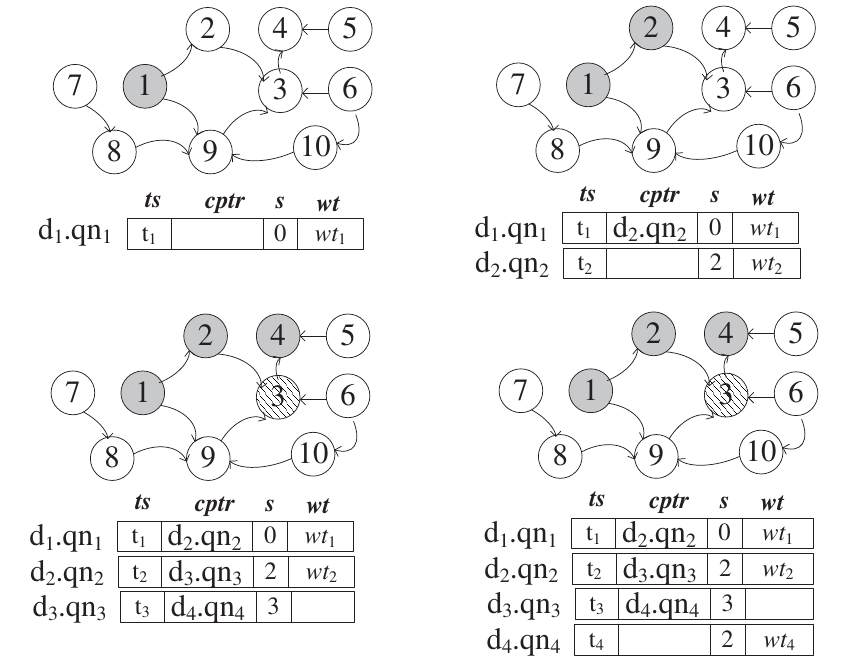}
	}
\caption{Evidence chain reasoning process examples}
\label{fig9ab}
\end{figure}

Figure ~\ref{fig:ECRP} illustrated the above-mentioned procedure. The elements in the data structure $D$ changes together with the evidences as shown in Figure~\ref{fig:ERNExample}.

\subsection{Timing Independent Evidence Chain Reasoning}

Whether the evidence chain reasoning algorithm can draw a correct conclusion depends on the correct time stamping of the evidences. In order to solve the inconsistent time-stamping problem, which is often a challenge, we propose a timing-independent evidence chain reasoning algorithm. The process is as follows:

Step 1. Initialization: fetch the next evidence $\varepsilon_i$ and map it to a vertex of the ERN of the given system, i.e. $map(\varepsilon_i)=n_i$. Then set $d_i.qn_i.ts$ as the timestamp $t_i$;

Step 2. Vertex checking: if $d_i.qn_i.s=1$ or $d_i.qn_i.s=3$, then $(d_i.qn_i.s)--$, $d_i.qn_i,wt=w_i$, $d_i.qn_i.ts=t_i$, then go back to step 1, otherwise go to step 3;

Step 3. Association analysis: the analysis methods vary based on different types of vertices of the ERN;

\begin{enumerate}
\item Case A: When $IN(n_i)=0$, $\varepsilon_i$ is the start point, set $d_i.qn_i.s=0$, and $d_i.qn_i.wt=w_i$;

\item Case B: When $IN(n_i)>0$, if we substitute all truth-value links into $l_i$ and $l_i$ is true, then $\varepsilon_i$ is an intermedia node. Thus we set $cptr=d_i.qn_i$ for all parent nodes that make $l_i$ true, and set $d_i.qn_i.s=2$, $d_i.qn_i.wt=w_i$;

\item Case C: $IN(n_i)>0$, if we substitute all truth-value link into $l_i$ and $l_i$ is false, then traverse all parent nodes of $n_i$ to search for one of the parent nodes $n_j$ which makes $l_i$ true when we increase a virtual record to $n_j$.
The virtual record generated in the reasoning process indicates that the evidence is not acquired by EventTracker, so the risk weights are not taken into account (see further explanation later in subsection \ref{power}). We further conduct association analysis on $n_j$, if case A is met, then set $d_j.qn_j.s=1$, $d_j.qn_j.cptr=d_i.qn_i$, $d_i.qn_i.s=2$, $d_i.qn_i.wt=w_i$; if case B is met, then set $d_j.qn_j.s=3$, $d_j.qn_j.cptr=d_i.qn_i$, $d_i.qn_i.s=2$, $d_i.qn_i.wt=w_i$; if case C is met, then $\varepsilon_i$ has no correlation with the other evidences. Thus we take $\varepsilon_i$ as a start node, and set $d_i.qn_i.s=0$, $d_i.qn_i.wt=w_i$.

\end{enumerate}

Step 4. Evidence chain generation: conduct breadth-first search from a ERN vertex, of which the in-degree is 0, and generate evidence chain based on the $cptr$ of the traversed vertices.

Figure~\ref{fig8} shows the approximate reasoning process, where vertex 3 is a virtual record.

\subsection{Time Complexity Analysis}

Let $N$ and $E$ be the number of vertices and edges/links in a given ERN. The first step of our evidence chain reasoning algorithm finishes in constant time. The time complexity of the second step is $O(E/N)$. And the time complexity of the third step is $O(N+E)$. Therefore, the overall time complexity of the evidence chain reasoning algorithm is $O(N+E)$.

For the time complexity of the `Timing Independent Evidence Chain Reasoning' algorithm, the first and the second step finish in constant time. Since the algorithm needs to determine the status of one of the parents node of a virtual node, the average time complexity of step 3 is $O((E/N)^2)$, and BFS time complexity is $O(N+E)$. For a directed complete graph, $E=N^2-N$, thus $N+E\geq(E/N)^2$. Hence, the time complexity of the `Timing Independent Evidence Chain Reasoning' algorithm is $O(N+E)$.

\subsection{Power of the evidence chain}\label{power}

The power of electronic evidence refers to the persuasive power of an evidence in proving the case. Given $W$ of an ERN defines the risk weight of every vertex, EventTracker adds up the risk weight of all evidences on an evidence chain and it will be used to evaluation the power of an evidence chain.

As mentioned earlier, the virtual record generated in the reasoning process indicates that the evidence is not acquired by the EventTracker. Therefore, its risk weight is not added into the overall weights $W$. Assume the sum of risk weights of all evidences in an evidence chain is $w$, and the sum of risk weights of all the virtual records on this evidence chain is $w'$, the weight of the evidence chain should be calculated as shown in formula \ref{equ1}. The bigger the value of $confidence$, the bigger the evidence chain power.

\begin{equation}\label{equ1}
confidence = \frac{w}{w+w'}
\end{equation}

\subsection{The application of an ERN in EventTracker}

System calls are generally used to change the states of an operation system, such as creating files, forking processes, changing registries, etc. Monitoring the changes of system calls can get many essential characteristics of program operations, and thus form the basis of dynamic security analysis. In an EventTracker, Virtual Machine Introspection technology (VMI) is used to monitor system calls in cloud computing hosts.

\begin{figure}
\centering
\includegraphics[width=0.2\textwidth]{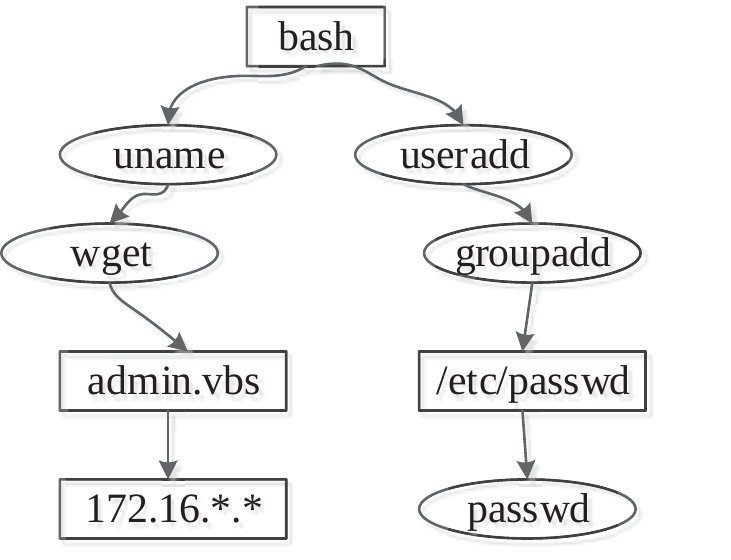}
\caption{Example of evidence reasoning network for system calls} \label{fig9}
\end{figure}

There are parameter dependencies among sequences of system calls generated by program operation. EventTracker maps system calls and their dependencies into an ERN model as the nodes and the links respectively. For example, Figure~\ref{fig9} illustrates an example of an Evidence Reasoning Network for system calls. First, an attacker uses \emph{useradd} and \emph{groupadd} commands to add system users and set permissions. This attacker then sets up passwords through \emph{passwd} commands. After checking the kernel version information with command \emph{uname}, "admin.vbs" script is downloaded from the remote host 172.16.*.* through FTP.

The goal of an EventTracker is to identify complex lateral movement attacks inside the cloud hosts. In the intrusion detection literature, an attack scenario (or attack pattern) is a sequence of explicit attack steps, which are logically linked and lead to an objective. When a set of correlation is received, if an attack scenario is detected, it will raise an intermediate attack alarm, which will prompt the system to capture the causal relationships among the evidences.

\subsection{Design and implementation of AlertCorrelator}

AlertCorrelator is located at the edge of a cloud computing environment. It correlates and analyzes alerts generated by multiple NIDS deployed in the cloud computing environment. NIDS is responsible for monitoring network traffic in real time, finding intrusion events and storing corresponding IP packets in a database. According to the design of ERN, we can see that AlertCorrelator does not tolerate false positives in alerts, but it can accept a certain degree of false negatives.

AlertCorrelator collects and preprocesses the alerts first. Every collected alert is coded and normalized into a standardized format. Additional information, such as timestamps and source address of the attacker are also added into the database. Events generated by different NIDS and related with the same attack are merged into a single alarm. Then these preprocessed alarms are added into the ERN as evidences in a chronological order. At last, timing independent evidence chain reasoning algorithm is called. In this phase, the ERN model determines whether the attack is either a successful attack or a non-relevant attack, i.e., an attack that does not lead to lateral movement attack.

For large scale networks, when new NIDS are added to the cloud boundary, augmenting an existing ERN can be done easily in an iterative fashion. This demonstrates the high scalability and flexibility of
AlertCorrelator.

\section{Experiment Evaluation}

To further verify the effectiveness of our proposed method, we developed our EventTracker prototype system using C language on Linux RedHat 7.3. We use the intrusion detection system Snort 2.4.3 as the attack detection component, and use Graphviz to visualize the evidence chain. We conduct our experiments on Shuguang servers (CPU 2.4G, MEM 8G). We report in this section two sets of experiments.

\subsection{Parameter Setup}

In order to evaluate the power of the evidence chain, we need to set an appropriate risk weights of ERN. Since the difficulty of utilizing each vulnerability in different situation is different, the probability of successful exploitation of each vulnerability is different. In our experiments, we assume that all the probability of successful exploitation of each vulnerability are equal, and the functional attributes of network nodes are also equal. As for the security impact of vulnerabilities, we classify the vulnerabilities with the impact scope into 9 categories and set each category with a certain weight, as shown in Table~\ref{tab2}.

\begin{table}
\centering
\caption{Weight of vulnerability impact}\label{tab1}
\begin{tabular}{cm{3.8cm}c}
\hline
\tiny Vulnerability & {\tiny impact scope} & \tiny Weight\\
\tiny  category & & \\
\hline
\tiny 1 & \tiny System administrators, managing system resources, system files, system processes, and other resources & \tiny 1.0\\
\hline
\tiny 2 & \tiny System administrator with partial permissions & \tiny 0.8\\
\hline
\tiny 3 & \tiny Permissions of any number of system ordinary users with more independent and private resources & \tiny 0.6\\
\hline
\tiny 4 & \tiny Permissions of a system ordinary user and partial permissions of other ordinary users & \tiny 0.5\\
\hline
\tiny 5 & \tiny Permissions of a system ordinary user created by the system initialization or created by the system administrator with its own private resources & \tiny 0.4\\
\hline
\tiny 6 & \tiny Partial permission of a ordinary user & \tiny 0.2\\
\hline
\tiny 7 & \tiny Remote visitors who can access network services, usually trusted visitors, who can interact with network service processes, scan system information, and so on & \tiny 0.1\\
\hline
\tiny 8 & \tiny Remote visitors who are connected to the target system at the physical layer, usually untrusted or firewalled visitors. & \tiny 0.0\\
\hline
\end{tabular}
\end{table}

\subsection{Reasoning Result of the Lincoln Dataset}

Experiment 1 uses the LLDOS1.0 and LLDOS2.0.2 data sets from MIT Lincoln Laboratories. The test bed producing this data set includes an external Internet environment simulated by 14 hosts, an internal network of 39 hosts, and a DMZ area consisting of 6 hosts, covering operating systems including Windows, Linux RedHat 5.0, SunOS 4.1.4, and Solaris 2.7. Attack detection component is used to produce an intranet data set of size 179 MB from each of the two LLDOS datasets.  The type and amount of evidences generated are shown in Fig~\ref{fig9ab}. Figure~\ref{fig10} shows the evidence chain deduced by EventTracker on LLDOS1.0.

This evidence chain indicates the attacker's five attack phases: the attacker first scans the entire network from host 202.77.162.213 to learn the target host's IP address range; then it executes Sadmind Overflow program with the ping option checked in order to verify on each host whether the Sadmind service is running. It then determines the final destination host (e.g. 172.16.115.20 in Figure~\ref{fig8}); next, the attacker uses the Solaris operating system's Sadmind vulnerability to implement a buffer overflow attack; after gaining the root privileges, the attacker installs the Mstream backdoor program through telnet and rpc; finally, the attacker uses this compromised slave to launch a DDoS attack to host 172.16.115.20/172.16.112.10/131.84.1.131. The evidence chain of the above-mentioned attacking process is shown in Figure~\ref{fig8}. This result matches the documen provided by MIT Lincoln, which proves the effectiveness of the proposed EventTracker. Since there is no virtual records, the $confidence$ is 100\% in this case.

The deducted evidence chain of LLDOS2.0.2 is shown in Figure~\ref{fig11}. This evidence chain also contains a complete attack sequence, and uses a springboard attack. Similar to LLDOS 1.0, the attacker also implemented an attack on host 202.77.162.213, and succeeded in obtaining the root privileges of hosts 172.16.115.20 and 172.16.112.50 that are using the Solaris operating system with Sadmind vulnerability. However, comparing to the former one, the LLDOS2.0.2 attack process is more complex and with latency. For example, the attacker did not use the "ICMP echo reply" which could be easily shielded. Instead, it uses the legitimate "DNS HINFO" query to obtain the target host (a DNS server). It is noted that the shadow node "FTP upload" in the evidence chain indicates that the attacker used ftp to upload the evidence of the Mstream backdoor program and attacking script. However, the attack detection component missed this event. To ensure the integrity of the evidence, EventTracker generated a virtual record. According to the weights set in Table~\ref{tab1}, the $confidence$ is reduced to 86\% accordingly. Then we use the timestamp of the alert of ``INFO TELNET access 119" to find the PCL. Two relations were found which is shown in  Figure~\ref{fig11}. This further proves the correctness of our reasoning algorithm.

\begin{figure}
\centering
    \subfigure[Types and quantity of evidences detected from LLDOS1.0 intranet  dataset]{
		\label{fig9a}
		\includegraphics[width=0.2\textwidth]{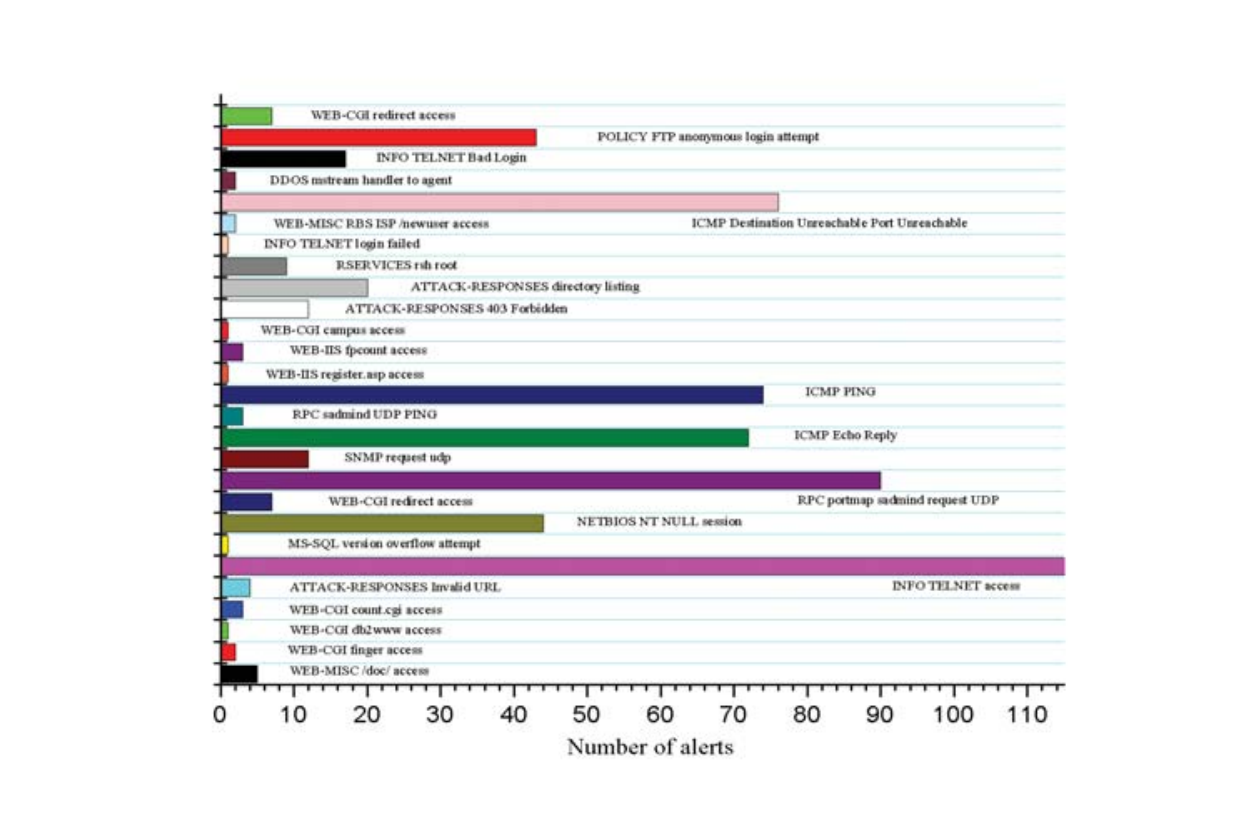}
	}
	\subfigure[Types and quantity of evidences detected from LLDOS2.0.2 intranet dataset]{
		\label{fig9b}
		\includegraphics[width=0.2\textwidth]{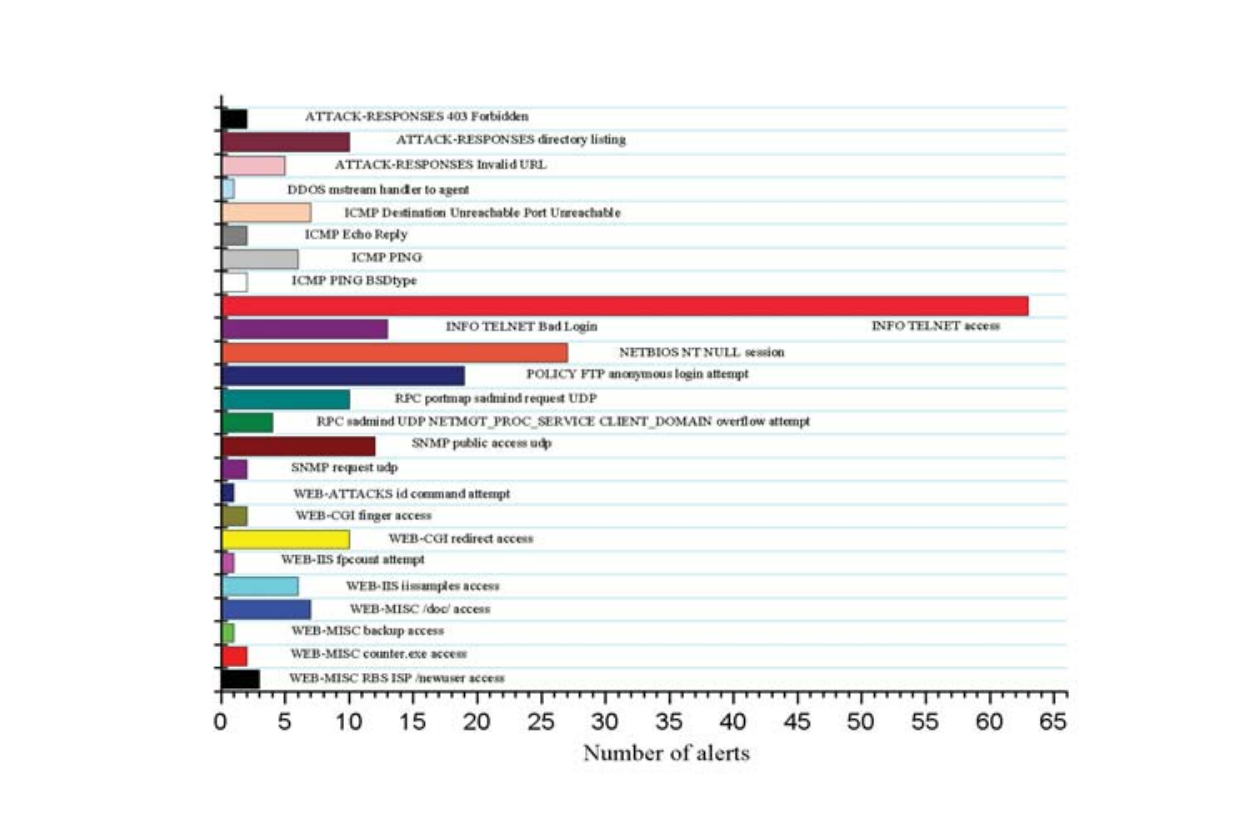}
	}
\caption{Categories and quantity of LLDOS dataset}
\label{fig9ab}
\end{figure}

\begin{figure}
\centering
\includegraphics[width=0.45\textwidth]{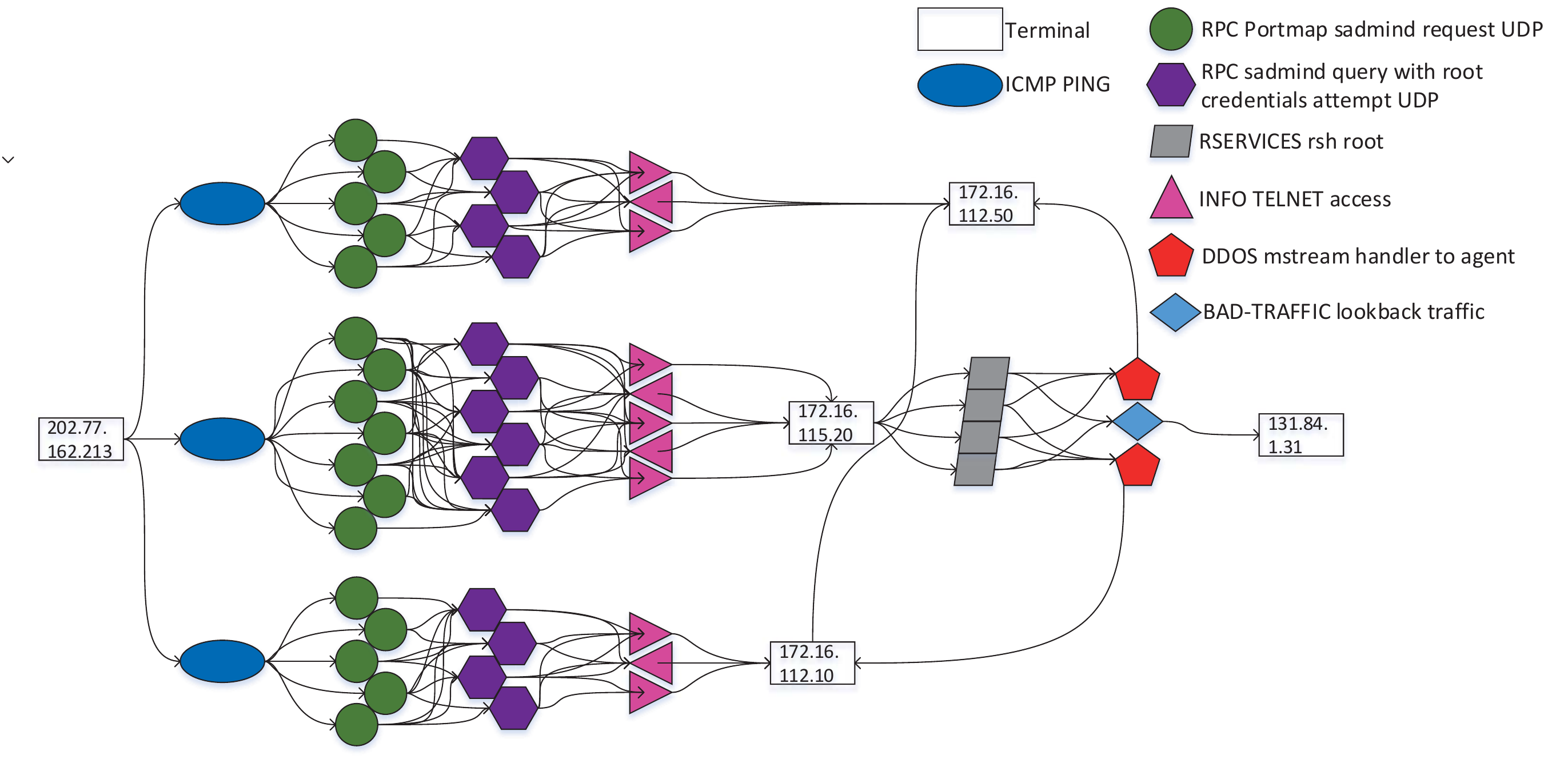}
\caption{Evidence Chain of LLDOS1.0 Intranet Dataset} \label{fig10}
\end{figure}

\begin{figure}
\centering
\includegraphics[width=0.45\textwidth]{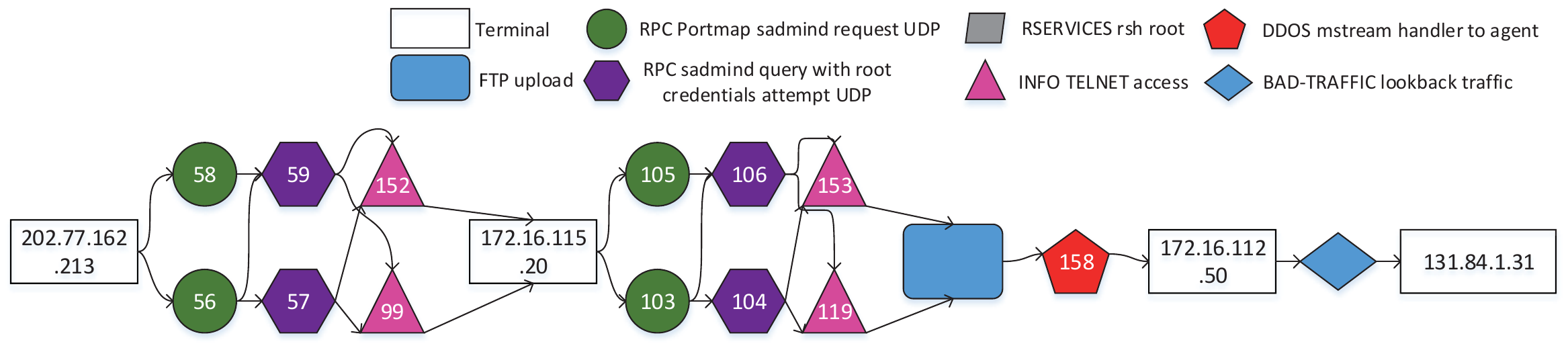}
\caption{Evidence chain of LLDOS2.0 data set.} \label{fig11}
\end{figure}

\subsection{Detection Result of Treasure Hunt Dataset}

\begin{figure}
\centering
\includegraphics[width=0.3\textwidth]{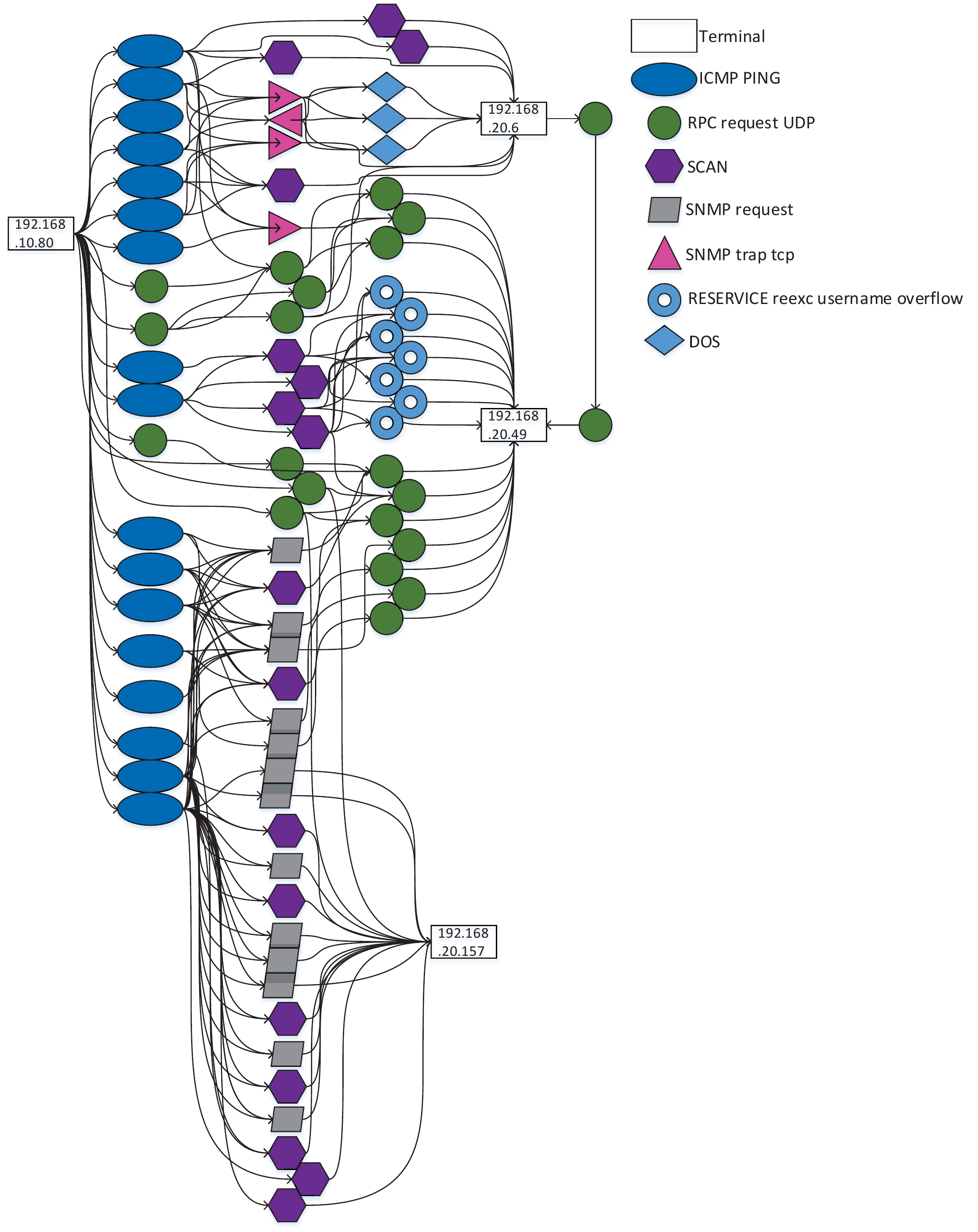}
\caption{Evidence Chain of the alpha sub-network.} \label{fig12}
\end{figure}

\begin{figure}
\centering
\includegraphics[width=0.3\textwidth]{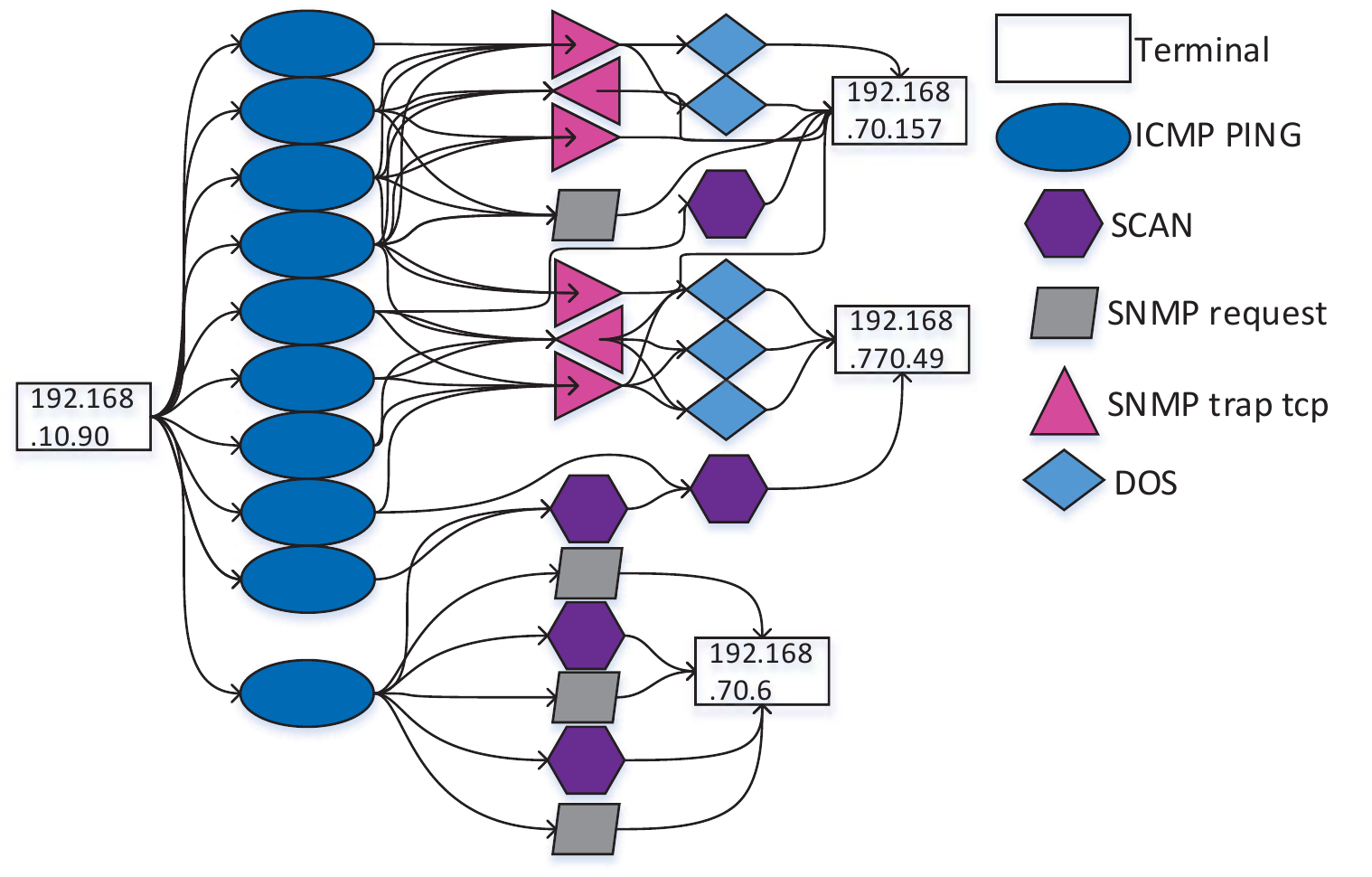}
\caption{Evidence Chain of the Omega sub-network} \label{fig13}
\end{figure}

The second experiment uses the Treasure Hunt dataset collected by the University of California Santa Barbara to guide students in the design of network offense and defense course. Its network topology is divided into three sub-networks: Alpha, Omega, and DMZ, which includs MySQL servers, event processing servers, file servers and WEB servers. Although the network topology is not complex, a wide range of attack methods are adopted in this network. Therefore, we believe that this dataset is suitable for the edge-cloud scenario functional testing of the proposed EventTracker. The corresponding evidence chains are shown in Figure~\ref{fig12} and Figure~\ref{fig13}, respectively, with $confidence$ interval of 91.7\% and 33.3\%.

\subsection{Performance Analysis}

Table~\ref{tab2} gives the time overhead of CloudSEC in handling the above sets of data. It is shown that the number of evidence chains CloudSEC could handle per second is around 100. According to \cite{19}, about 10-20000 security events per day on average are detected by each detection point. CloudSEC greatly exceeds the current actual processing requirements under the conditions of real-time reasoning lateral movement, this frontage is suitable for deploying in the edge-cloud environment.

\begin{table}
\centering
\caption{The time overhead of CloudSEC}\label{tab2}
\begin{tabular}{m{1.1cm}ccm{1cm}m{1cm}m{1.5cm}}
\hline
\multicolumn{3}{c}{\tiny Data Sets} & \tiny Intrusion Events & \tiny Process Time (sec) & \tiny Average Processing Speed (one event per secnod)\\
\hline \\
\multirow{4}*{\tiny MIT/LL 2000} & \multirow{2}*{\tiny LLDOS 1.0} & \tiny DMZ & \tiny 2498 & \tiny 15.9806 & \multirow{4}*{\tiny 78.8505} \\
\cline{3-5} \\
 &   & \tiny Inside&\tiny 905&\tiny 16.7467& \\
 \cline{2-5} \\
 & \multirow{2}*{\tiny LLDOS 2.0.2} & \tiny DMZ&\tiny 1125&\tiny 16.5234& \\
 \cline{3-5} \\
 &   & \tiny Inside&\tiny 624&\tiny 16.0881& \\
 \hline
\multirow{2}*{\tiny Treasure Hunt} & \multicolumn{2}{c}{\tiny Alpha}  &\tiny 732&\tiny 15.8154& \multirow{2}*{\tiny 119.8598} \\
\cline{2-5} \\
 & \multicolumn{2}{c}{\tiny Omega}  &\tiny 732&\tiny 15.8154&  \\
\hline
\end{tabular}
\end{table}

\section{Conclusion}

In this paper, we propose a new method to track events for Lateral Movement Detection. The concept of vulnerability correlation is introduced. Methods on how to construct an ERN based on the vulnerability knowledge and network environment information is provided. Then two lateral movement reasoning algorithms based on the constructed ERN are presented. The proposed CloudSEC provides a strong guarantee for the rapid and effective evidence investigation as well as real-time attack detection, this advantage is more suitable for those complex edge-cloud computing environments, for example, the services based on the collaboration between the edge artificial intelligence and the cloud computing. Experiments using various real network datasets proves the correctness of the proposed approach. Theoretical analysis concludes that both chain reasoning algorithms achieve linear time complexity. In the future, we aim at improving the performance on ERN generation which further improves the overall CloudSEC efficiency, at the same time multi-type of lateral movement tricks will be evaluated.


\section*{Acknowledgment}
This work is supported by the National Natural Science Foundation of China under NO.
61572153, NO.
61702220, NO.
61702223, and NO.
U1636215. And the National Key research and Development Plan (Grant No.
2018YFB0803504).

\vspace{12pt}

\end{document}